# Crack-tip stress evaluation of multi-scale Griffith crack subjected to tensile loading by using peridynamics

Xiao-Wei Jiang, Hai Wang*

School of Aeronautics and Astronautics, Shanghai Jiao Tong University, Shanghai 200240, China

**ABSTRAT**

Crack-tip stress evaluation has always been a problem in the frame of classical elasticity theory. Peridynamics has been shown to have great advantages in dealing with crack problems. In the present study, we present a peridynamic crack-tip stress evaluation method for multi-scale Griffith crack subject to tensile loading. The bond-based peridynamics is used to calculate the displacement field. Non-local deformation gradient definition from non-ordinary state-based peridynamics is used for stress calculation. Besides, a scale factor is introduced for evaluating crack-tip stress of multi-scale Griffith crack. Numerical results compared with Eringen's results show that this peridynamic crack-tip stress evaluation method is valid for multi-scale cracks, and with the change of distance of material points, the evaluated crack-tip stress tends to be stable.

**Keywords:** Crack-tip stress; Peridynamics; Non-local; Multi-scale

## 1. Introduction

Crack-tip stress evaluation has always been a problem in the frame of classical elasticity theory. According to the results of linear elastic fracture mechanics, the crack-tip stress will increase to infinite. In order to solve this crack-tip stress singularity problem, Eringen *et al.* [1977] proposed the non-local elasticity and presented results of finite crack-tip stress. The non-local elasticity abandons the assumption that stress at a point is only related to the strain at that point. Instead, it uses a non-local stress-strain relationship, where stress at a point is related to the strain both at that point and its surrounding horizon.

Non-local elasticity has been widely used for crack-tip stress evaluation. Zhou *et al.* [1999] determined the state of stress in a plate with a Griffith crack subject to the anti-plane shear by using the non-local theory. Tovo *et al.* [2007] addressed the problem of stress singularities at the tip of sharp V-notches by means of a non-local implicit gradient approach. Ghosh *et al.* [2013] developed an integral type non-local continuum model for epoxy from phonon dispersion data, which can be used to regularize the stress field at crack tips and molecular defect cores. Jamia *et al.* [2014] considered the problem of a mixed-mode crack embedded in an infinite medium made of a functionally graded magneto-electro-elastic material (FGMEEM) with the crack surfaces subjected to magneto-electro-mechanical loadings and non-local theory of elasticity is applied to obtain the governing magneto-electro-elastic equations.

Silling [2000] derived the peridynamic theory (PD) for analysis of discontinuous problems. Peridynamics has been shown to have great advantages in the simulation of crack propagation. Kilic and Madenci [2009] employed the peridynamic theory to predict crack growth patterns in quenched glass plates. Silling *et al.* [2010] proposed a condition for the emergence of a discontinuity in an elastic peridynamic body, resulting in a material stability condition for crack nucleation. Ghajari *et al.* [2014] proposed a new material model for the dynamic crack propagation analysis of anisotropic materials within the framework of bond-based peridynamic theory. Lee and Hong [2016] presented peridynamic simulation on crack branching and curving in a pre-exisitng center-notched brittle polymer. De Meo *et al.* [2016] presented a numerical Multiphysics peridynamic framework for the modelling of adsorbed-hydrogen stress-corrosion cracking (SCC). The peridynamic theory uses integration, instead of differentiation, to compute the force on a material point. The material point within a finite horizon can interact with each other, therefore the peridynamic theory can be categorized as a non-local theory [Silling, 2000; Breitenfeld *et al.*, 2014].

Since peridynamics uses the similar non-local concept as Eringen's non-local elasticity does, and peridynamics brings out great advantageous in dealing with crack problems, a motivation is to use peridynamics to evaluate the crack-tip stress as Eringen's non-local elasticity could do. In the present paper, we focus our attention on the peridynamic evaluation of crack-tip stress for multi-scale Griffith crack subjected to tensile loading. Firstly, in Section 2, we present a peridynamic crack-tip stress evaluation method for multi-scale Griffith crack subjected to tensile loading. Then, in Section 3, we illustrate that this peridynamic crack-tip stress evaluation method is valid for micro-scale Griffith crack under tensile loading. Finally, in Section 4, we extend this peridynamic crack-tip stress evaluation method to macro-scale Griffith crack under tensile loading and show that the peridynamic evaluation result of crack-tip stress tends to be stable.

**2. Peridynamic crack-tip stress evaluation method for multi-scale Griffith crack**

A peridynamic crack-tip stress evaluation method for multi-scale Griffith crack subjected to tensile loading is presented here. This evaluation method has three steps:
(1) Calculate the displacement field using bond-based peridynamics;
(2) Calculate the crack-tip stress based on the non-local definition of deformation gradient from non-ordinary state-based peridynamics (NOSB PD) [Silling et al., 2007];
(3) Evaluate the crack-tip stress of multi-scale Griffith crack through multiplying the result from step (2) by a scale factor.
The scale factor in step (3) will be given in Section 2.3, and the reason for introducing this factor will be discussed in Section 3.3.

2.1 Bond-based peridynamic theory

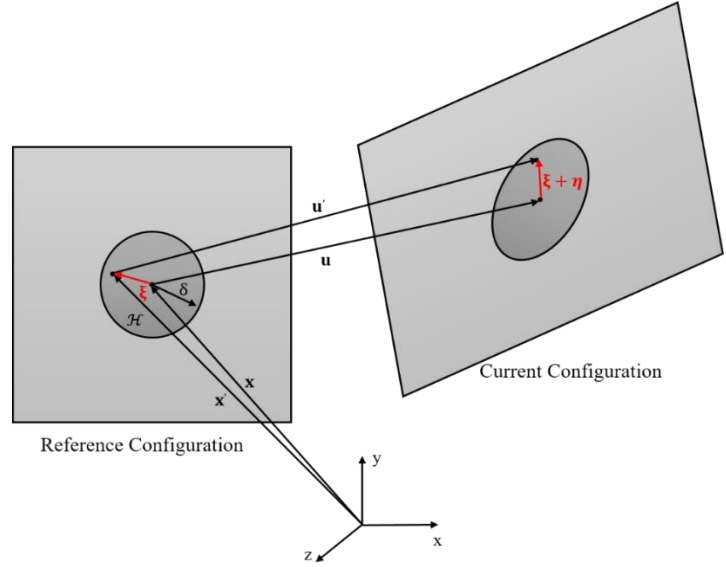

**Fig. 1.** Peridynamic notations

In the present paper, bond-based peridynamics proposed by Silling [2000] was used to calculate the displacement field. As shown by Silling and Askari [2005], the acceleration of any material point at $\mathbf{x}$ in in reference configuration at time $t$ is found from

$$\rho(\mathbf{x})\ddot{\mathbf{u}}(\mathbf{x},t) = \int_H \mathbf{f}(\mathbf{u}(\mathbf{x}',t) - \mathbf{u}(\mathbf{x},t), \mathbf{x}' - \mathbf{x})dV_{\mathbf{x}'} + \mathbf{b}(\mathbf{x},t) \qquad (1)$$

where $\mathbf{u}$ and $\mathbf{u}'$ are displacements at $\mathbf{x}$ and $\mathbf{x}'$, respectively, $H$ is the horizon zone, as shown in Fig. 1, and $\boldsymbol{\xi} = \mathbf{x}' - \mathbf{x}$, $\boldsymbol{\eta} = \mathbf{u}' - \mathbf{u}$. $\rho(\mathbf{x})$ is the density, and $\mathbf{f}$ is a pairwise force function defined as

$$\mathbf{f}(\boldsymbol{\eta}, \boldsymbol{\xi}) = \frac{\partial w}{\partial \boldsymbol{\eta}}(\boldsymbol{\eta}, \boldsymbol{\xi}) \quad \forall \boldsymbol{\xi}, \boldsymbol{\eta} \qquad (2)$$

where $w$ is a scalar named as "micropotential" and a material is said to be microelastic if Eq. (2) is satisfied. Besides, there exists a scalar-valued function

$$\hat{w}(y, \boldsymbol{\xi}) = w(\boldsymbol{\eta}, \boldsymbol{\xi}) \quad \forall \boldsymbol{\xi}, \boldsymbol{\eta} \quad y = |\boldsymbol{\eta} + \boldsymbol{\xi}| \qquad (3)$$

Combining Eq. (2) and (3) and differentiating the latter with respect to the components of $\boldsymbol{\eta}$ leads to

$$\mathbf{f}(\mathbf{\eta},\mathbf{\xi}) = \frac{\mathbf{\xi}+\mathbf{\eta}}{|\mathbf{\xi}+\mathbf{\eta}|} f(|\mathbf{\xi}+\mathbf{\eta}|,\mathbf{\xi}) \quad \forall \mathbf{\xi},\mathbf{\eta} \tag{4}$$

From Eq. (2) and (3) to Eq. (4), the detailed derivation was omitted by Silling and Askari [2005]. For better understanding of the direction of pairwise force, here we supplement the detailed derivation:

$$\mathbf{f}(\mathbf{\eta},\mathbf{\xi}) = \frac{\partial w}{\partial \mathbf{\eta}}(\mathbf{\eta},\mathbf{\xi}) = \frac{\partial \hat{w}}{\partial \mathbf{\eta}}(y,\mathbf{\xi}) = \frac{\partial \hat{w}(y,\mathbf{\xi})}{\partial y} \frac{\partial y}{\partial \mathbf{\eta}} \tag{5}$$

$$\frac{\partial \hat{w}(y,\mathbf{\xi})}{\partial y} = f(y,\mathbf{\xi}) = f(|\mathbf{\xi}+\mathbf{\eta}|,\mathbf{\xi}) \tag{6}$$

$$y = |\mathbf{\xi}+\mathbf{\eta}| = \sqrt{(x_1' - x_1 + u_1' - u_1)^2 + (x_2' - x_2 + u_2' - u_2)^2} \tag{7}$$

$$\frac{\partial y}{\partial \mathbf{\eta}} = \begin{bmatrix} \frac{\partial y}{\partial \eta_1} \\ \frac{\partial y}{\partial \eta_2} \end{bmatrix} = \begin{bmatrix} \frac{\partial y}{\partial (u_1' - u_1)} \\ \frac{\partial y}{\partial (u_2' - u_2)} \end{bmatrix} = \begin{bmatrix} \frac{1}{2}\left\{(x_1' - x_1 + u_1' - u_1)^2 + (x_2' - x_2 + u_2' - u_2)^2\right\}^{-1/2} \times 2(x_1' - x_1 + u_1' - u_1) \\ \frac{1}{2}\left\{(x_1' - x_1 + u_1' - u_1)^2 + (x_2' - x_2 + u_2' - u_2)^2\right\}^{-1/2} \times 2(x_2' - x_2 + u_2' - u_2) \end{bmatrix}$$

$$= \begin{bmatrix} \dfrac{x_1' - x_1 + u_1' - u_1}{\sqrt{(x_1' - x_1 + u_1' - u_1)^2 + (x_2' - x_2 + u_2' - u_2)^2}} \\ \dfrac{x_2' - x_2 + u_2' - u_2}{\sqrt{(x_1' - x_1 + u_1' - u_1)^2 + (x_2' - x_2 + u_2' - u_2)^2}} \end{bmatrix} = \begin{bmatrix} \dfrac{\xi_1 + \eta_1}{|\mathbf{\xi}+\mathbf{\eta}|} \\ \dfrac{\xi_2 + \eta_2}{|\mathbf{\xi}+\mathbf{\eta}|} \end{bmatrix} = \frac{\mathbf{\xi}+\mathbf{\eta}}{|\mathbf{\xi}+\mathbf{\eta}|}$$

$$\tag{8}$$

From Eq. (6) and (8), we can easily get Eq. (4). And Eq. (8) also illustrates that the direction of pairwise force is parallel to the deformed bond.

For homogeneous elastic isotropic material, the scalar bond force $f$ in Eq. (4) only depends on the bond stretch, defined by

$$s = \frac{|\mathbf{\xi}+\mathbf{\eta}| - |\mathbf{\xi}|}{|\mathbf{\xi}|} \tag{9}$$

$$f = cs \tag{10}$$

Comparing the peridynamic strain energy density to the classical theory of elasticity strain energy density under isotropic extension, the spring constant $c$ in Eq. (10) could be expressed as

$$c = \frac{18\kappa}{\pi\delta^4} \tag{11}$$

where $\kappa$ is the bulk modulus and $\delta$ is the radius of the horizon, $H$.

As the peridynamic governing equation (1) is in dynamic form, the adaptive dynamic relaxation (ADR) method proposed by Kilic and Madenci [2010] is used for static problem.

According to the ADR method, Eq. (1) at the $n^{th}$ iteration can be rewritten as

$$\ddot{\mathbf{U}}^n(\mathbf{X},t^n) + c^n\dot{\mathbf{U}}^n(\mathbf{X},t^n) = \mathbf{D}^{-1}\mathbf{F}^n(\mathbf{U}^n,\mathbf{U}'^n,\mathbf{X},\mathbf{X}') \tag{12}$$

where $\mathbf{D}$ is the fictitious diagonal density matrix and $c$ is the damping coefficient which can be expressed by

$$c^n = 2\sqrt{((\mathbf{U}^n)^T {}^1\mathbf{K}^n\mathbf{U}^n)/((\mathbf{U}^n)^T\mathbf{U}^n)} \tag{13}$$

in which ${}^1\mathbf{K}^n$ is the diagonal "local" stiffness matrix, which is given as

$${}^1K_{ii}^n = -(F_i^n/\lambda_{ii} - F_i^{n-1}/\lambda_{ii})/(\Delta t \dot{u}_i^{n-1/2}) \tag{14}$$

where $F_i^n$ is the value of force vector $\mathbf{F}^n$ at material point $\mathbf{x}$, which includes both the peridynamic force state vector and external forces. And $\lambda_{ii}$ is the diagonal elements of $\mathbf{D}$ which should be large enough for numerical convergence.

By utilizing central-difference explicit integration, displacements and velocities for the next time step can be obtained.

$$\dot{\mathbf{U}}^{n+1/2} = \frac{((2-c^n\Delta t)\dot{\mathbf{U}}^{n-1/2} + 2\Delta t\mathbf{D}^{-1}\mathbf{F}^n)}{(2+c^n\Delta t)} \tag{15}$$

and

$$\dot{\mathbf{U}}^{n+1} = \mathbf{U}^n + \Delta t\dot{\mathbf{U}}^{n+1/2} \tag{16}$$

To start the iteration process, we assume that $\mathbf{U}^0 \neq 0$ and $\dot{\mathbf{U}}^0 = 0$, so the integration can be started by

$$\dot{\mathbf{U}}^{1/2} = \frac{\Delta t\mathbf{D}^{-1}\mathbf{F}^0}{2} \tag{17}$$

## 2.2 Stress calculation in non-ordinary state-based peridynamics

As the stress concept is not obvious in bond-based peridynamics, some definitions from NOSB PD is used here to get the calculated value of crack-tip stress. The most important definition used is the non-local definition of deformation gradient **F** [Warren et al., 2009]

$$\mathbf{F}(\mathbf{x}_k) = \left[\sum_{i=1}^{n} w\langle \mathbf{x}_i - \mathbf{x}_k\rangle(\underline{\mathbf{Y}}\langle \mathbf{x}_i - \mathbf{x}_k\rangle \otimes (\mathbf{x}_i - \mathbf{x}_k)V_i\right] \cdot \mathbf{K}^{-1}(\mathbf{x}_k) \tag{18}$$

$$\mathbf{K}(\mathbf{x}_k) = \sum_{i=1}^{n} w\langle \mathbf{x}_i - \mathbf{x}_k\rangle((\mathbf{x}_i - \mathbf{x}_k) \otimes (\mathbf{x}_i - \mathbf{x}_k))V_i \tag{19}$$

where **K** is the shape tensor and $w\langle \mathbf{x}' - \mathbf{x}\rangle$ is the influence function defined by

$$w\langle \mathbf{x}' - \mathbf{x}\rangle = \frac{\delta}{|\mathbf{x}' - \mathbf{x}|} \tag{20}$$

After the non-local deformation gradient is got, stress can be easily calculated by using the knowledge from nonlinear continuum mechanics, as shown by Fan and Li [2016]

$$\mathbf{E} = \frac{1}{2}(\mathbf{F}^T\mathbf{F} - \mathbf{I}) \tag{21}$$

where **E** is the green strain tensor and **I** is the identity matrix. Green strain is energy conjugate with second Piola-Kirchhoff stress tensor $\boldsymbol{\sigma}^{PK2}$

$$\boldsymbol{\sigma}^{PK2} = \mathbf{C}:\mathbf{E} \tag{22}$$

where **C** is the stiffness tensor, and for 2D isotropic problems, **C** is a $4\times4\times4\times4$ tensor.

$$\mathbf{C} = \begin{bmatrix} \dfrac{E}{1-v^2} & 0 & 0 & \dfrac{E}{1+v} \\ 0 & \dfrac{E\mu}{1-v^2} & 0 & 0 \\ 0 & 0 & \dfrac{E\mu}{1-v^2} & 0 \\ \dfrac{E}{1+v} & 0 & 0 & \dfrac{E}{1-v^2} \end{bmatrix} \tag{23}$$

where $E$ is elastic modulus and $v$ is Poisson' ratio.

Normally, we need the Cauchy stress tensor $\boldsymbol{\sigma}$, and

$$\boldsymbol{\sigma}^{PK2} = J\mathbf{F}^{-1} \cdot \boldsymbol{\sigma} \cdot \mathbf{F}^{-T} \quad (24)$$

Transfer Eq. (24) and we can get

$$\boldsymbol{\sigma} = \frac{1}{J}\mathbf{F}\boldsymbol{\sigma}^{PK2}\mathbf{F}^T \quad (25)$$

2.3 The scale factor for evaluating crack-tip stress of multi-scale Griffith crack

Here we directly give this scale factor. The reason for the usage of the scale factor will be discussed in Section 3.3.

$$\sigma_{crt} = (d/d_0)^{1/2} \sigma_{crt}^{PD} \quad (26)$$

where $\sigma_{crt}$ is the evaluated crack-tip stress and $\sigma_{crt}^{PD}$ is the calculated crack-tip stress got from Eq. (25). $d$ is the distance of the material point of peridynamic model, and $d_0$ is the atomic distance as described by Eringen *et al.* [1977].

## 3. Crack-tip stress evaluation of micro-scale Griffith crack subjected to tensile loading

In the current section, firstly we set the distance of material point in peridynamic model equals the atomic distance, and the scale factor in Eq. (26) is 1. We will show that at this circumstance, the evaluated crack-tip stress is valid compared with Eringen's results [1977]. Secondly, we vary the distance of material point in peridynamic model, and we will see that by using the scale factor in Eq. (26), a stable and valid crack-tip stress can also be evaluated. Finally, a few discussions will be given about the usage of this scale factor.

3.1 $d = d_0$

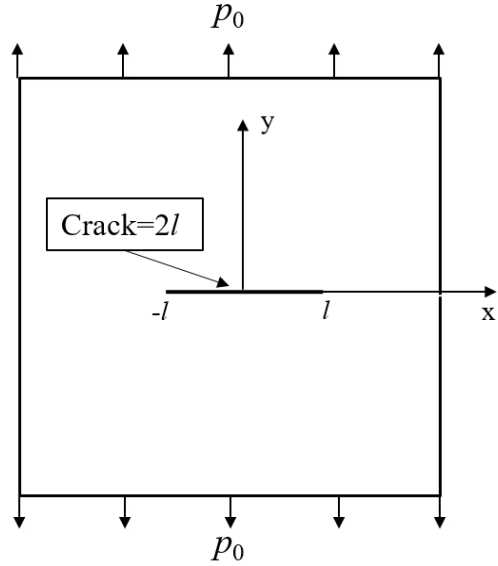

**Fig. 2.** A typical Griffith crack problem

For a typical Griffith crack subject to tensile loading, as shown in Fig. 2, Eringen [1977] gave the finite crack-tip stress distribution for $2l/d_0 = 20, 40$ and $100$. And an analytical crack-tip stress was given

$$\sigma_{crt} \approx 0.73(2l/d_0)^{1/2} p_0 \qquad (27)$$

where $2l$ is the crack length, $d_0$ is the atomic distance, and $p_0$ is the external load pressure.

Setting the distance of material point $d$ equals atomic distance $d_0$, we use peridynamics to calculate the stress field of a micro-scale Griffith crack under tensile loading. The resulting $\sigma_{yy}/p_0$ distribution is shown in Fig. 3.

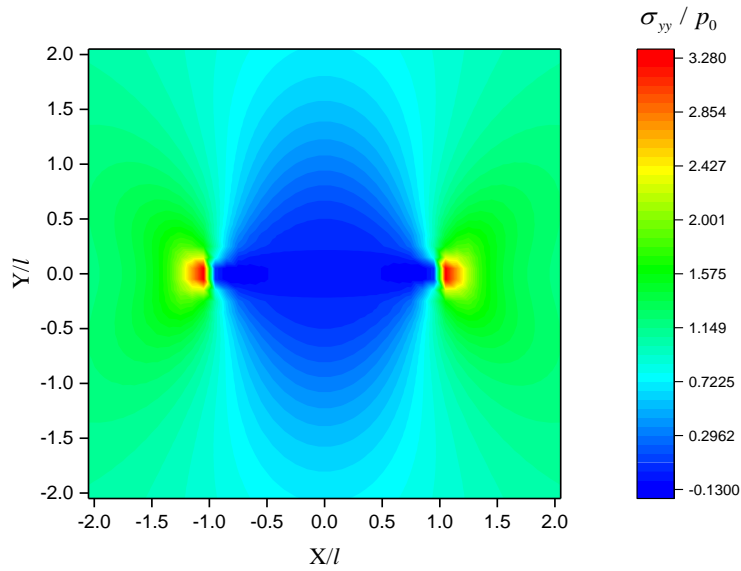

**Fig. 3.** Peridynamic $\sigma_{yy}/p_0$ distribution of Griffith crack subject to tensile loading, $2l/d_0 = 20$

Comparing with Eringen's results, the crack-tip $\sigma_{yy}/p_0$ distribution for $2l/d_0 = 20$, 40, and 100 are shown in Fig. 4. It can be seen from the figure that the peridynamic crack-tip stress distribution fits the Eringen's results well. The crack-tip stress distribution becomes more and more sharp with the increase of the crack length, and the stress concentration also becomes more and more serious at the crack-tip.

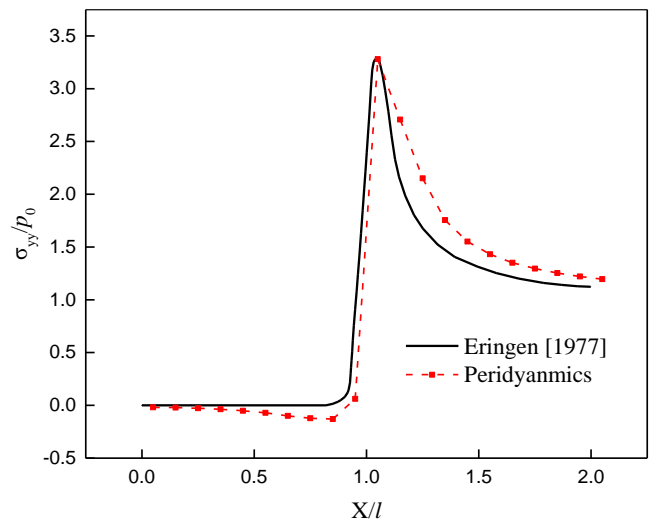

(a) $2l/d_0 = 20$

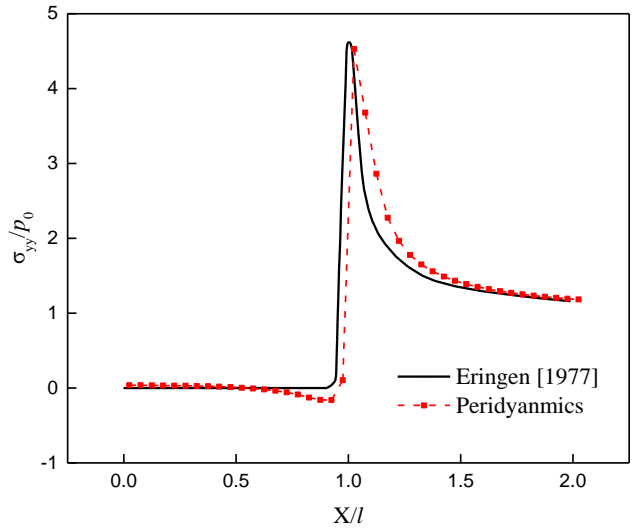

(b) $2l/d_0 = 40$

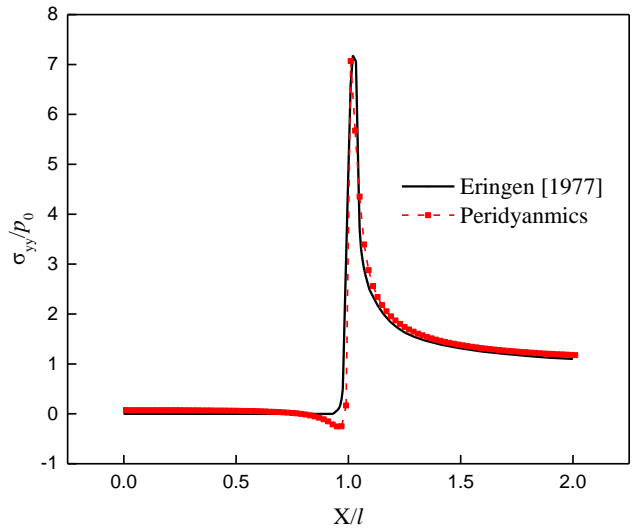

(c) $2l/d_0 = 100$

**Fig. 4.** Crack-tip $\sigma_{yy}/p_0$ distribution of Griffith crack subject to tensile loading, $(d/d_0)^{1/2} = 1$

Crack-tip stress evaluation results of micro-scale Griffith crack subject to tensile loading are given in Table 1. Six kinds of micro-scale Griffith crack with different $2l/d_0$ are given. Two kinds of materials are discussed. From the table, we can see that compared with Eringen's results, the evaluated peridynamic crack-tip results are within the acceptable error when $(d/d_0)^{1/2} = 1$. For different material, with the same $2l/d_0$, crack-tip stress is nearly the same. It may be worth noting that the atomic distance $d_0$ for different material is different. So for the same crack length, the crack-tip stress of different material is different. This phenomenon seems obvious, as different material has different elastic modulus. But we should note that this difference discussed here is from the point of atomic distance of material, which may reveals that the crack-tip stress evaluation problem is an interesting multi-scale problem, especially for macro-scale crack.

**Table 1**

Crack-tip stress evaluation of micro-scale Griffith crack subject to tensile loading, $p_0 = 200\text{MPa}$, $(d/d_0)^{1/2} = 1$

| $2l/d_0$ | Eringen (MPa) | Peridynamics (MPa) | | | |
|---|---|---|---|---|---|
| | | Steel ($d_0$=2.48Å) | Error | diamond ($d_0$=1.54Å) | Error |
| 20 | 652.9 | 655.9 | 0.45% | 687.3 | 5.26% |
| 40 | 923.4 | 906.1 | -1.87% | 928.9 | 0.60% |
| 60 | 1130.9 | 1101.1 | -2.64% | 1119.9 | -0.97% |
| 80 | 1305.9 | 1266.8 | -2.99% | 1283.1 | -1.74% |
| 100 | 1460.0 | 1413.4 | -3.19% | 1428.1 | -2.18% |
| 120 | 1599.3 | 1546.4 | -3.31% | 1559.8 | -2.47% |

3.2 Changing the distance of material point $d$

In the previous section, we have shown that when setting $(d/d_0)^{1/2} = 1$, the evaluated crack-tip stress is acceptable. Here we want to show that with the change of $d$, the evaluated crack-tip stress are also acceptable and trend to be stable.

As presented in Table 2, by using Eq. (26), the crack-tip stress of a Griffith crack ($2l/d_0 = 100$) subjected to tensile loading is evaluated through multiplying the peridynamic result by a scale factor $(d/d_0)^{1/2}$. Comparing the results with Eringen's analytical result in Table 1, we can see that after multiplying the scale factor $(d/d_0)^{1/2}$

, the evaluated crack-tip stresses are all within acceptable error. Besides, from Fig. 5, we can see that with the decrease of $d/d_0$, the error trends to be stable.

**Table 2**

Crack-tip stress evaluation of micro-scale Griffith crack subject to tensile loading, $p_0 = 200\text{MPa}$, $2l/d_0 = 100$, $d_0 = 2.48\text{Å}$

| $d/d_0$ | PD result (MPa) | Evaluate result (MPa) | Error |
|---|---|---|---|
| 10.00 | 484.5 | 1532.1 | 4.94% |
| 5.00 | 655.9 | 1466.6 | 0.45% |
| 2.50 | 906.1 | 1432.7 | -1.87% |
| 1.67 | 1101.1 | 1421.5 | -2.64% |
| 1.25 | 1266.8 | 1416.3 | -2.99% |
| 1.00 | 1413.4 | 1413.4 | -3.19% |

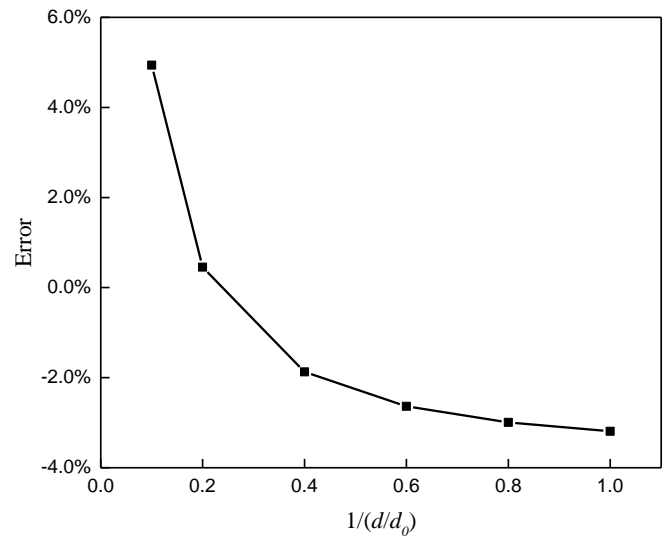

**Fig. 5.** Trend of the error with varying $d/d_0$ for a micro-scale Griffith crack subject to tensile loading, $p_0 = 200\text{MPa}$, $2l/d_0 = 100$, $d_0 = 2.48\text{Å}$

3.3 Discussion on the usage of scale factor $(d/d_0)^{1/2}$

From the previous results, we have shown that with the introducing of a scale factor $(d/d_0)^{1/2}$ in Eq. (26), the evaluated crack-tip stress of micro-scale Griffith crack subject to tensile loading is valid at any $d/d_0$. This is especially important for crack-tip stress evaluation of macro-scale Griffith crack, since for a macro-scale crack, we are unable to set the distance of material point $d$ to be equal to atomic distance $d_0$.

The above discussion is purely from the point of developing a valid peridynamic evaluation method of crack-tip stress. From another perspective, this introducing of a scale factor $(d/d_0)^{1/2}$ may also reveals that the crack-tip stress evaluation problem could be an interesting multi-scale problem, and this may be the reason why this problem can't be solved in the frame of classical elasticity.

**4. Crack-tip stress evaluation of macro-scale Griffith crack subjected to tensile loading**

The peridynamic crack-tip stress evaluation method presented in Section 2 has been proved to be valid for micro-scale Griffith crack subjected to tensile loading in Section 3. In the current section, we want to extend this method to macro-scale Griffith crack subjected to tensile loading.

For a macro-scale Griffith crack with crack length $2l = 1mm$, and atomic distance of material $d_0 = 2.48\text{Å}$, the evaluated crack-tip stresses are presented in Table 3 with the changing of $d/d_0$. From Fig. 6, we can easily see that with the decrease of $d/d_0$, the evaluated crack-tip stress trends to be stable.

**Table 3**

Crack-tip stress evaluation of macro-scale Griffith crack subject to tensile loading, $p_0 = 200\text{MPa}$, $2l = 1mm$, $d_0 = 2.48\text{Å}$

| $d/d_0$ | PD result (MPa) | Evaluate result (MPa) |
|---|---|---|
| 4.03E+08 | 484.5 | 9.73E+06 |
| 2.02E+08 | 655.9 | 9.31E+06 |
| 1.01E+08 | 906.1 | 9.10E+06 |
| 6.72E+07 | 1101.1 | 9.03E+06 |
| 5.04E+07 | 1266.8 | 8.99E+06 |
| 4.03E+07 | 1413.4 | 8.98E+06 |

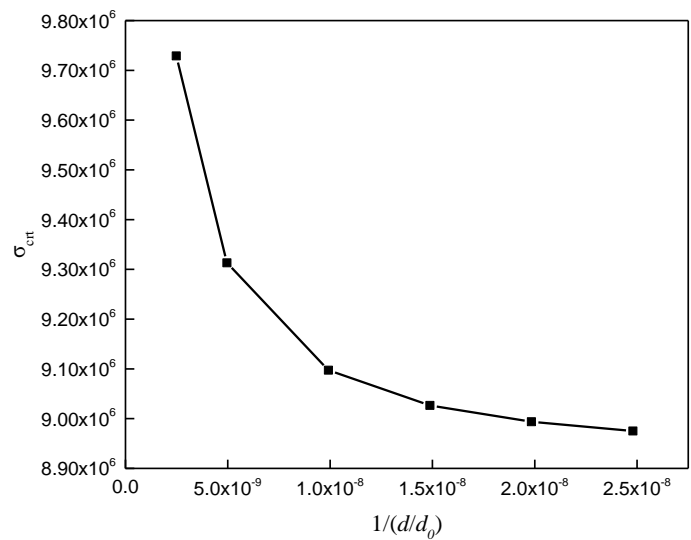

**Fig. 6.** Trend of the evaluated crack-tip stress with varying $d/d_0$ for a macro-scale Griffith crack subjected to tensile loading, $p_0 = 200\text{MPa}$, $2l = 1mm$, $d_0 = 2.48\text{Å}$

## 5 Conclusion

Crack-tip stress of multi-scale Griffith crack subjected to tensile loading was evaluated by using peridynamics. A peridynamic crack-tip stress evaluation method was presented. Bond-based peridynamics was used to calculate the displacement field. The non-local deformation gradient definition from NOSB PD was used for the calculation of crack-tip stress. A scale factor $(d/d_0)^{1/2}$ was introduced for evaluating crack-tip stress. Numerical results illustrate that this peridynamic evaluation method is valid for both micro-scale and macro-scale Griffith crack subjected to tensile loading. With the changing of $d/d_0$, the evaluated crack-tip stress tends to be stable.

## Acknowledgments

The authors are indebted to Prof. Hui-Shen Shen of Shanghai Jiao Tong University for his considerable support.

## References

Breitenfeld, M. S., Geubelle, P. H., Weckner, O., and Silling, S. A. [2014] "Non-ordinary state-based peridynamic analysis of stationary crack problems," *Computer Methods in Applied Mechanics and Engineering* **272**, 233-250.

De Meo, D., Diyaroglu, C., Zhu, N., Oterkus, E., and Siddiq, M. A. [2016] "Modelling of stress-corrosion cracking by using peridynamics," *International Journal of Hydrogen Energy*, **41**(15), 6593-6609.

Eringen, A. C., Speziale, C. G., and Kim, B. S. [1977] "Crack-tip problem in non-local elasticity," *Journal of the Mechanics and Physics of Solids* **25**(5), 339-355.

Fan, H., and Li, S. [2016] "Parallel peridynamics–SPH simulation of explosion induced soil fragmentation by using OpenMP," *Computational Particle Mechanics* 1-13.

Ghajari, M., Iannucci, L., and Curtis, P. [2014] "A peridynamic material model for the analysis of dynamic crack propagation in orthotropic media," *Computer Methods in Applied Mechanics and Engineering* **276**, 431-452.

Ghosh, S., Kumar, A., Sundararaghavan, V., and Waas, A. M. [2013] "Non-local modeling of epoxy using an atomistically-informed kernel," *International Journal of Solids and Structures* **50**(19), 2837-2845.

Warren, T. L., Silling, S. A., Askari, A., Weckner, O., Epton, M. A., and Xu, J. [2009] "A non-ordinary state-based peridynamic method to model solid material deformation and fracture," *International Journal of Solids and Structures* **46**(5), 1186-1195.

Lee, J., and Hong, J. W. [2016] "Dynamic crack branching and curving in brittle polymers," *International Journal of Solids and Structures* **100**, 332-340.

Silling, S. A. [2000] "Reformulation of elasticity theory for discontinuities and long-range forces," *Journal of the Mechanics and Physics of Solids* **48**(1), 175-209.

Silling, S. A., and Askari, E. [2005] "A meshfree method based on the peridynamic model of solid mechanics," *Computers and Structures* **83**(17), 1526-1535.

Silling, S. A., Epton, M., Weckner, O., Xu, J., and Askari, E. [2007] "Peridynamic states and constitutive modeling," *Journal of Elasticity* **88**(2), 151-184.

Jamia, N., El-Borgi, S., Rekik, M., and Usman, S. [2014] "Investigation of the behavior of a mixed-mode crack in a functionally graded magneto–electro-elastic material by use of the non-local theory," *Theoretical and Applied Fracture Mechanics* **74**, 126-142.

Kilic, B., and Madenci, E. [2009] "Prediction of crack paths in a quenched glass plate by using peridynamic theory," *International journal of fracture* **156**(2), 165-177.

Kilic, B., Madenci, E. [2010] "An adaptive dynamic relaxation method for quasi-static simulations using the peridynamic theory," *Theoretical and Applied Fracture Mechanics* **53**(3), 194-204.

Kohlhoff, S., Gumbsch, P., and Fischmeister, H. F. [1991] "Crack propagation in bcc crystals studied with a combined finite-element and atomistic model," *Philosophical Magazine A*, **64**(4), 851-878.

Silling, S. A., Weckner, O., Askari, E., and Bobaru, F. [2010] "Crack nucleation in a peridynamic solid," *International Journal of Fracture* **162**(1-2), 219-227.

Tovo, R., and Livieri, P. [2007] "An implicit gradient application to fatigue of sharp notches and weldments," *Engineering Fracture Mechanics* **74**(4), 515-526.